\documentclass[preprint,12pt]{elsarticle}

\usepackage{lineno,hyperref}

\usepackage[lmargin=.75in,rmargin=.75in,tmargin=1.in,bmargin=1in]{geometry}
\usepackage{amsmath}
\usepackage{amsthm}
\usepackage{amsfonts}
\usepackage{amssymb}
\usepackage{float}
\usepackage{graphicx}
\usepackage{booktabs}
\usepackage{caption}
\usepackage{subcaption}
\usepackage[utf8]{inputenc}
\usepackage{algorithm}
\usepackage{enumitem}
\usepackage{algorithmicx, algpseudocode}
\usepackage{natbib}

\newtheorem{proposition}{Proposition}
\newtheorem{remark}{Remark}
\newtheorem{assumption}{Assumption}

\newcommand{\R}{\mathbb{R}}
\newcommand{\E}{\mathbb{E}}

\newcommand{\N}{\mathcal{N}}
\graphicspath{{figs/}}

\journal{Inverse Problems}

\title{A generalized ensemble-marginalized Kalman filter for semilinear time-dependent PDEs with noisy boundary conditions}

\begin{document}

\begin{frontmatter}

\author[kfupm1,kfupm2]{Zaid Sawlan\corref{corrauthor}}
\cortext[corrauthor]{Corresponding author.}
\ead{zaid.sawlan@kfupm.edu.sa}

\address[kfupm1]{Department of Mathematics, King Fahd University of Petroleum \& Minerals, Dhahran, Saudi Arabia}
\address[kfupm2]{IRC for Refining \& Advanced Chemicals, King Fahd University of Petroleum \& Minerals, Dhahran, Saudi Arabia}

\begin{abstract}
Uncertainty in boundary data is a major source of bias in PDE-constrained inverse problems, yet it is often neglected by treating measured boundary records as exact inputs. The ensemble-marginalized Kalman filter (EnMKF) removes this bias for linear parabolic problems with noisy boundaries by modeling boundary values as a linear Gaussian control, estimating them with a Kalman filter, and marginalizing their uncertainty from the prediction step. We extend this construction to semilinear time-dependent PDEs by exploiting time discretizations in which the boundary enters the discrete dynamics affinely, such as IMEX schemes that treat the linear part implicitly and the nonlinear reaction explicitly. For such discretizations the boundary uncertainty can still be marginalized one step at a time, yielding a deterministic covariance inflation term without any additional sampling. For second-order-in-time problems, the same per-step marginalization fails because the sensitivity of the discrete velocity to boundary errors scales like the inverse time step; we introduce an error-state EnMKF that propagates temporally correlated boundary-error realizations according to the innovations-form dynamics of the control Kalman filter. Parameter-dependent observation operators are treated by augmenting the ensemble with predicted observations and projecting the marginalization term into observation space. Numerical experiments on contaminant transport with nonlinear decay and on nonlinear cable vibration demonstrate reduced boundary-induced parameter bias, avoidance of ensemble collapse at small ensemble sizes, and improved state and flux estimates relative to standard ensemble Kalman filters.
\end{abstract}

\begin{keyword}
data assimilation \sep ensemble Kalman filter \sep noisy boundary
conditions \sep joint state-parameter estimation \sep semilinear PDEs
\end{keyword}

\end{frontmatter}

\section{Introduction}
\label{sec:intro}

Data assimilation for systems governed by partial differential
equations usually assumes that the boundary conditions are known.  In
many applications they are not.  The boundary values are measured by
sensors, and those measurements carry noise of the same order as the
noise in the observations used for the assimilation itself.  Treating
the measured boundary records as exact inputs produces biased estimates
of physical parameters, a problem documented in detail for the thermal
characterization of building walls \citep{Iglesias2018IJHMT,
Ruggeri2017}.  The bias does not vanish as more interior data are
collected, because the error enters through the model rather than
through the likelihood.

For linear parabolic problems, \citet{Iglesias2018IP} proposed the
ensemble-marginalized Kalman filter (EnMKF) where the boundary values $u_k$ are modeled as a linear Gaussian state-space system driven by an autoregressive evolution and observed with noise. Therefore, classical Kalman filter \citep{Kalman1960} provides the filtering distribution $p(u_k \mid z_{1:k})$ in closed form. The PDE state $T_k$, conditional on the physical parameters $\theta$, is linear in $u_k$, so the boundary uncertainty can be integrated out of the prediction step analytically.  The result is a deterministic inflation of the prediction covariance by the expected value of $B_\theta P^u_{k|k} B_\theta'$ over the parameter ensemble, where $B_\theta$ is the discrete control operator and $P^u_{k|k}$ is the filtered covariance of the boundary values. It is considered inflation in the sense of an additive covariance contribution;
in contrast to the ad hoc inflation factors common in ensemble
filtering \citep{AndersonAnderson1999}, this term is derived in closed
form from the control model and has no free tuning parameter.
Compared with an EnKF
that samples the boundary values or ignores their noise, the marginalized filter avoided
ensemble collapse without artificial inflation and reached the mean-field posterior with half the ensemble size or less.

Viewed in the framework of sequential Monte Carlo, the construction is a
Rao-Blackwellization of the ensemble Kalman filter: the linear Gaussian boundary
control is integrated out in closed form, while the remaining nonlinear
state and parameter uncertainties are treated with an ensemble update.
The error-state variant developed below preserves this principle, since
it propagates the boundary error with the exact dynamics of the control
Kalman filter rather than resampling it independently at each step.

Using an ensemble Kalman framework over fully Bayesian
samplers is common because particle filters degenerate in
high-dimensional state spaces \citep{Snyder2008}, Markov chain Monte Carlo requires repeated full model runs \citep{Stuart2010,
KaipioSomersalo2005}, and the EnKF family
\citep{Evensen1994} offers a tractable
compromise whose behavior is well documented
\citep{Evensen2009, LawStuart2015, KatzfussStroudWikle2016}.  Joint
state-parameter estimation by state augmentation has a long record in
hydrology and geophysics \citep{Moradkhani2005}, and its known
weaknesses, ensemble collapse and the resulting need for covariance
inflation \citep{AndersonAnderson1999, SchillingsStuart2017}, are
precisely the issues
the marginalization was shown to remove without tuning.  Simplified
lumped models remain common in the building-physics application
itself \citep{Biddulph2014}, increasing the value of methods that
make PDE-constrained assimilation cheap enough for routine use.
Boundary-condition uncertainty is also a recognized error source well
beyond building physics, for example, in flood inundation forecasting
\citep{Pappenberger2006}.

The restriction to linear dynamics is the main obstacle to wider use of
this idea.  Reaction terms, radiative boundary exchange, and
concentration-dependent decay all destroy global linearity, yet they
appear in exactly the applications where boundary data are noisy:
building physics, subsurface transport, and structural dynamics.
Since then, related work has approached the joint state-parameter
problem through Rao-Blackwellized and marginalized particle filters
\citep{SchonGustafsson2005, DoucetRB2000, IqbalClaudel2024}, through
hybrid and low-rank ensemble constructions \citep{Bach2023}, and
through regularization of iterative ensemble methods for the parameter-only
problem \citep{IglesiasLawStuart2013, Iglesias2016, DeSimon2018}.  The
EnMKF shares the
marginalization principle of these methods, but differs in the treatment
of the remaining nonlinear block.  A Rao-Blackwellized particle filter
keeps a particle approximation of $(T_k, \theta)$ and a Kalman filter
for $u_k$ per particle, and the particle weights degenerate when the
state $T_k$ is high dimensional \citep{Snyder2008}; the EnMKF instead
represents $(T_k, \theta)$ by an ensemble Kalman update, which scales to
the discretized PDE states considered here at the cost of the
Gaussian-analysis approximation. None of these methods address the
specific structure of noisy boundary forcing in nonlinear PDE dynamics.

This paper generalizes the EnMKF to semilinear time-dependent PDEs.
The key observation is elementary but useful: the marginalization step
never required linear dynamics.  It requires only that the boundary
data enter the one-step discrete map affinely, and implicit-explicit
(IMEX) time discretizations \citep{AscherRuuthWetton1995} of semilinear
problems have exactly this property.  The conditional distribution of
the predicted state given the parameters and the previous state remains
Gaussian with computable moments, and the prediction step of the EnMKF
carries over verbatim with the linear propagator replaced by a
nonlinear map.  This affine dependence is a property of the time
discretization rather than of the physics, and whether it survives as
the order of the time derivative increases separates the paper into two
contributions.

The first contribution is this parabolic filter itself: under the affine
property just described, the boundary uncertainty is marginalized out of
the prediction step exactly (Proposition~\ref{prop:marg}), which carries
the EnMKF from linear to semilinear parabolic problems.  One
extension keeps this parabolic filter usable in the applications that
motivate it, without altering its structure.  Observation operators of
the form $h_\theta(T) = D_\theta H T$, linear in the state but dependent
on the parameters, arise, for instance, when a diffusive flux is scaled by an unknown
diffusivity; the ensemble is augmented with the predicted
observations and the marginalization term is projected onto the
observation space analytically, which is consistent and avoids the
instability we observed with innovation-rescaling schemes under wide
priors.

The second contribution concerns second-order-in-time dynamics, where
per-step marginalization fails.  The sensitivity of the discrete
velocity to the current boundary value scales like $2/\Delta t$, so
treating the filtered boundary uncertainty as independent across steps
inflates the velocity variance by orders of magnitude when the boundary
error is temporally persistent, and the inflation worsens without bound
as the time step is refined.  We resolve this with an error-state
variant of the filter: each ensemble member carries a realization of the
boundary error that evolves with the exact innovations-form error
dynamics of the control Kalman filter, so the temporal correlation of
the boundary error is represented without approximation.  The diagnosis
and this construction, rather than the use of an error-state formulation
as such, are the contribution; together they carry the
ensemble-marginalization principle into second-order-in-time problems, a
setting the per-step marginalization underlying the original filter
cannot reach.

The methods are evaluated on two synthetic-data applications, one for
each regime: contaminant transport with second-order decay and a noisy
mass-transfer boundary, a parabolic problem for which the marginalization
is exact;
and the vibration of a cable on a nonlinear elastic foundation driven by
noisy support-motion records, a second-order-in-time problem for which
per-step marginalization is inadequate and the error-state variant is
required.  In each case we compare the generalized EnMKF
with a modified EnKF that samples the boundary values from their
filtered distribution, and with a noise-blind EnKF that plugs the raw
measurements into the boundary.

The paper is organized as follows.  Section~\ref{sec:formulation} sets
up the semilinear state-space formulation and the control model.
Section~\ref{sec:method} develops the parabolic filter, the extension
to parameter-dependent observations, the error-state variant for
second-order-in-time dynamics, and the baseline filters.
Section~\ref{sec:examples} reports the numerical experiments.
Section~\ref{sec:conclusions} summarizes the findings and discusses
limitations.

\section{Semilinear state-space formulation}
\label{sec:formulation}

Consider the semilinear initial-boundary value problem on a bounded
domain $D \subset \R^d$ and time interval $[0, t_N]$,
\begin{equation}
\label{eq:pde}
\begin{cases}
\dfrac{\partial T}{\partial t} + L_\theta T = f_\theta(T),
  & x \in D,\; t \in (0, t_N],\\[4pt]
\mathcal{B}_\theta T = u(x,t), & x \in \partial D,\; t \in (0, t_N],\\[2pt]
T(x,0) = T_0(x), & x \in D,
\end{cases}
\end{equation}
where $L_\theta$ is a linear elliptic operator with coefficients
collected in the parameter vector $\theta \in \R^p$, the reaction term
$f_\theta$ is a (possibly nonlinear) function of the state acting
pointwise, and $\mathcal{B}_\theta$ is a boundary operator of the form
$\mathcal{B}_\theta T = \alpha T + \beta\, \partial T / \partial \nu$,
which covers Dirichlet ($\beta = 0$), Neumann ($\alpha = 0$), and Robin
boundary conditions.  The boundary data $u$ are not known.  They are
observed by sensors, and the same formulation accommodates uncertain
source amplitudes by stacking them into $u$.  Second-order-in-time
problems are written as first-order systems in the pair (displacement,
velocity) and fit the same template; the wave example of
Section~\ref{sec:ex3} gives the details.

We work under the following standing assumption.

\begin{assumption}
\label{ass:reg}
For every $\theta$ in the parameter set, the operator $L_\theta$ is
uniformly elliptic with bounded coefficients, the reaction term
$f_\theta$ is Lipschitz in $T$ with at most polynomial growth, and
$\mathcal{B}_\theta$ defines a well-posed boundary-value problem for
\eqref{eq:pde}.  The spatial discretization is stable and consistent,
and the IMEX time stepping is stable on the time scales considered.
\end{assumption}

These conditions are stated at the level the filtering construction
requires, and we do not prove well-posedness of \eqref{eq:pde} or
convergence of the filters as the discretization is refined.  Their role
is to guarantee that the discrete model introduced next is well defined
and that its one-step map is Lipschitz, which is all the marginalization
uses.  The continuum problem \eqref{eq:pde} accordingly serves as
motivation, and the results are formulated at the discrete level.

Discretizing \eqref{eq:pde} in space (finite differences or finite
elements) and in time with an IMEX scheme that treats $L_\theta$
implicitly and $f_\theta$ explicitly \citep{AscherRuuthWetton1995}
yields the discrete model
\begin{equation}
\label{eq:state}
T_k = g_\theta(T_{k-1}) + B_\theta u_k + w_k, \qquad
g_\theta(T) := A_\theta\bigl(T + \Delta t\, F_\theta(T)\bigr),
\end{equation}
where $T_k \in \R^n$ is the state vector at time $t_k$, $A_\theta =
(I + \Delta t\, L_h(\theta))^{-1}$ is the implicit propagator, $F_\theta$
is the discretized reaction term, $B_\theta = \Delta
t\, A_\theta
R_h(\theta)$ collects the boundary (and source) couplings, $u_k \in
\R^\ell$ is the control vector, and $w_k \sim \N(0, W_k)$ is model
noise.  The essential structural property of \eqref{eq:state} is that
$u_k$ enters affinely even though $g_\theta$ is nonlinear in both the
state and the parameters.  Stabilized explicit treatments of the
nonlinearity, such as Patankar-type updates \citep{BurchardPatankar2003}, preserve this
property.  The class of models the method covers is therefore the
discrete-time systems of the form \eqref{eq:state} with $g_\theta$
possibly nonlinear in the state and the parameters but with affine
control entry; the IMEX discretization of \eqref{eq:pde} is one route to
this structure rather than the only one, and the exact marginalization
result of Section~\ref{sec:enmkf} is a statement about \eqref{eq:state}
that does not refer to the continuum equation.

Partial observations of the state are collected at the same times,
\begin{equation}
\label{eq:obs}
y_k = h_\theta(T_k) + v_k, \qquad v_k \sim \N(0, V_k),
\end{equation}
with $y_k \in \R^m$.  We first take $h_\theta(T) = H T$ linear and
parameter independent, and return to parameter-dependent operators in
Section~\ref{sec:condlinear}.

The control vector follows the linear Gaussian companion model of
\citet{Iglesias2018IP},
\begin{equation}
\label{eq:control}
s_k = F^u\, s_{k-1} + q_k, \qquad u_k = H^u s_k, \qquad z_k = u_k + c_k,
\end{equation}
where $s_k$ is the companion state of an autoregressive control model,
$H^u$ selects the current control $u_k$ from it, $q_k \sim \N(0, Q_k)$,
$c_k \sim \N(0, C_k)$, and $z_k$ are the noisy
boundary measurements.  When the true boundary evolution is unknown,
the autoregressive model is taken as a random walk AR(1), $u_k =
u_{k-1} + q_k$ (so $s_k = u_k$, $F^u = I$, $H^u = I$), or a
random-increment AR(2), $u_k = 2u_{k-1} - u_{k-2}
+ q_k$ (so $s_k = (u_k, u_{k-1})$ and $F^u$, $H^u$ are the corresponding
companion matrices).  The AR(1) model is used for the reservoir
concentration of Section~\ref{sec:ex1}, and the AR(2) model, which tracks
smooth signals with locally linear trends, for the support motion of
Section~\ref{sec:ex3}.
The noises $w_k$, $v_k$, $q_k$, $c_k$ are mutually independent and
independent of $T_0$ and $\theta$.

The goal is to characterize $p(T_k, \theta \mid z_{1:k}, y_{1:k})$
sequentially.  Since \eqref{eq:control} is linear, Gaussian, and
autonomous from the state, the filtering distribution $p(u_k \mid
z_{1:k}) = \N(u_{k|k}, P^u_{k|k})$ is available exactly from a Kalman
filter, and the joint posterior factorizes through
\begin{equation}
\label{eq:factorization}
p(T_k, \theta, u_k \mid z_{1:k}, y_{1:k-1})
 = p(T_k, \theta \mid u_k, z_{1:k-1}, y_{1:k-1})\,
   p(u_k \mid z_{1:k}),
\end{equation}
which is the Rao-Blackwellized structure the filter exploits
\citep{DoucetRB2000, SchonGustafsson2005}.

\section{Methodology}
\label{sec:method}

The methodology follows the two contributions of the introduction.
Sections~\ref{sec:kfu}--\ref{sec:condlinear} develop the parabolic
filter: the control Kalman filter and its exact error dynamics, the
exact marginalization for semilinear dynamics, and the extension to
parameter-dependent observations.  Section~\ref{sec:errorstate} turns to
second-order-in-time dynamics, where per-step marginalization fails, and
develops the error-state variant.  Section~\ref{sec:baselines} records
the baseline filters used in the comparisons.

\subsection{Kalman filter for the controls and its error dynamics}
\label{sec:kfu}

The Kalman filter applied to \eqref{eq:control} produces, at each step,
the filtered mean $u_{k|k}$ and covariance $P^u_{k|k}$ through the
standard prediction and analysis recursions.  Two additional quantities from the
same recursion are needed later.  Recall the companion state $s_k$
and the selection matrix $H^u$ of \eqref{eq:control}, and write
$K^u_k$ for the Kalman gain.  The filtered
error $\varepsilon_k := s_k - s_{k|k}$ satisfies the exact recursion
\begin{equation}
\label{eq:errordyn}
\varepsilon_k = \Phi_k\, \varepsilon_{k-1} + \eta_k, \qquad
\Phi_k = (I - K^u_k H^u) F, \qquad
\eta_k \sim \N(0, \Xi_k),
\end{equation}
with $\Xi_k = (I - K^u_k H^u) Q_k (I - K^u_k H^u)' + K^u_k C_k
(K^u_k)'$.  Equation \eqref{eq:errordyn} is the innovations form of the
filter error; it involves no approximation.  Its stationary marginal is
the filtered covariance, and its temporal correlation is governed by
$\Phi_k$, which is close to the identity whenever the filter trusts the
autoregressive model more than the measurements.  Persistent, slowly
decorrelating boundary errors are therefore the rule rather than the
exception for smooth boundary signals.

\subsection{Exact marginalization for semilinear dynamics}

The first result states that the boundary marginalization of the EnMKF
does not depend on linearity of the dynamics.

\begin{proposition}
\label{prop:marg}
Let the state evolve according to \eqref{eq:state} and let $p(u_k \mid
z_{1:k}) = \N(u_{k|k}, P^u_{k|k})$.  Assume $u_k$ is conditionally
independent of $T_{k-1}$ given $z_{1:k}$.  Then, for any $\theta$ and
$T_{k-1}$,
\begin{equation}
\label{eq:margpred}
T_k \mid \theta, T_{k-1}, z_{1:k} \;\sim\;
\N\!\bigl(g_\theta(T_{k-1}) + B_\theta u_{k|k},\;
B_\theta P^u_{k|k} B_\theta' + W_k\bigr).
\end{equation}
\end{proposition}

\begin{proof}
Conditional on $\theta$ and $T_{k-1}$, the map $u_k \mapsto
g_\theta(T_{k-1}) + B_\theta u_k$ is affine, $u_k$ is Gaussian with the
stated moments, and $w_k$ is an independent Gaussian.  An affine
transformation of a Gaussian plus an independent Gaussian is Gaussian,
with mean and covariance as in \eqref{eq:margpred}.
\end{proof}

The conditional independence assumption is inherited from
\citet{Iglesias2018IP}, and we state precisely what it costs.

\begin{remark}
\label{rem:condindep}
After the first assimilation step the filtered boundary error and the
state error are correlated, because the same boundary realizations have
driven both.  The exact joint filtering distribution therefore does not
factor as in \eqref{eq:factorization} beyond $k = 1$, even when the
dynamics and the observations are linear and Gaussian.
Proposition~\ref{prop:marg} discards this correlation, so the EnMKF is
an approximate filter irrespective of linearity, and the term $B_\theta
P^u_{k|k} B_\theta'$ is the boundary contribution that would be exact if
$u_k$ were independent of $T_{k-1}$ given $z_{1:k}$.  The induced error
is small when that correlation is small, which holds for strongly
dissipative dynamics, for assimilation windows short relative to the
mixing time of the discrete semigroup, and more generally whenever
$\|B_\theta\|$ is bounded uniformly in $\Delta t$ so that past boundary
influence decays.  Section~\ref{sec:errorstate} shows that for
second-order-in-time dynamics none of these conditions holds, and
removes the assumption for the boundary error.
\end{remark}

\subsection{The generalized EnMKF}
\label{sec:enmkf}

Let $X_k = (\theta, T_k)$ denote the augmented vector and let
$\{(\theta^i, T^i_{k-1|k-1})\}_{i=1}^M$ be the analysis ensemble at
time $t_{k-1}$, with static parameters so that $\theta^i_{|k-1} =
\theta^i$.  The prediction ensemble is the conditional mean of
Proposition~\ref{prop:marg} evaluated member by member,
\begin{equation}
\label{eq:predmean}
T^i_{k|k-1} = g_{\theta^i}\bigl(T^i_{k-1|k-1}\bigr)
 + B_{\theta^i}\, u_{k|k},
\end{equation}
which is deterministic: neither the boundary values nor the model noise
is sampled.  Their uncertainty enters the gain through the prediction
covariance, derived next.

By Proposition~\ref{prop:marg}, conditional on $\zeta = (\theta,
T_{k-1})$ the augmented vector $X_k$ has mean $\bigl(\theta,\,
g_\theta(T_{k-1}) + B_\theta u_{k|k}\bigr)$ and a covariance whose only
nonzero block is $\Sigma_\theta := B_\theta P^u_{k|k} B_\theta' + W_k$ in
the state coordinates.  The prediction covariance of the ensemble is the
sample covariance of the predicted means \eqref{eq:predmean} plus the
ensemble average of these conditional covariances,
\begin{equation}
\label{eq:predcov}
\mathcal{P}_{k|k-1} \approx
\frac{1}{M-1} \sum_{i=1}^{M}
\bigl(X^i_{k|k-1} - \bar X_{k|k-1}\bigr)
\bigl(X^i_{k|k-1} - \bar X_{k|k-1}\bigr)'
 + \frac{1}{M} \sum_{i=1}^{M}
\begin{bmatrix} \mathbf{0} & \mathbf{0} \\
\mathbf{0} & B_{\theta^i} P^u_{k|k} B_{\theta^i}' + W_k \end{bmatrix}.
\end{equation}
Formula \eqref{eq:predcov} is the marginalized prediction covariance of
the EnMKF \citep{Iglesias2018IP}, here obtained for the semilinear map
$g_\theta$: the boundary and model uncertainty enter the gain
deterministically, through the second sum, without any sampling of the
boundary values or the model noise.

The analysis step is the stochastic EnKF update with perturbed
observations \citep{BurgersEnKF1998, Evensen2009}: with $\mathcal{H} =
[\mathbf{0}\;\; H]$,
\begin{equation}
\label{eq:analysis}
K_k = \mathcal{P}_{k|k-1} \mathcal{H}'
\bigl(\mathcal{H} \mathcal{P}_{k|k-1} \mathcal{H}' + V_k\bigr)^{-1},
\qquad
X^i_{k|k} = X^i_{k|k-1} + K_k\bigl(y_k + v^i_k
 - \mathcal{H} X^i_{k|k-1}\bigr),
\end{equation}
with $v^i_k \sim \N(0, V_k)$.  We report the analysis in its
stochastic, perturbed-observation form throughout; since the
marginalization modifies only the prediction covariance, a
deterministic square-root update would serve equally well, and the
choice does not bear on the conclusions.  Positive physical parameters
are carried
in logarithmic coordinates inside $X_k$, and when the prior support is
bounded the updated log-parameters are projected back onto the prior
box; without this safeguard the parameter ensemble can drift into
regimes where the explicit nonlinear term is unstable before the data
become informative (Section~\ref{sec:ex3}).  In implementations, only
the products $\mathcal{P}\mathcal{H}'$ and
$\mathcal{H}\mathcal{P}\mathcal{H}'$ are formed, so the full
$(p+n)\times(p+n)$ covariance is never assembled.  The per-step cost of
the marginalization is the cost of computing $B_{\theta^i} v$ for the
$\ell$ columns of $P^u_{k|k}$, i.e.\ $\ell$ extra implicit solves per
member, reusing the factorization of the propagation step.  Algorithm
\ref{alg:enmkf} summarizes the filter.

\begin{algorithm}[t]
\caption{Generalized EnMKF for semilinear PDEs (deterministic
inflation)}
\label{alg:enmkf}
\begin{enumerate}\setlength{\itemsep}{1pt}
\item draw $\theta^i \sim \rho_\Theta$ and $T^i_0 \sim \N(T_0, P_0)$,
      $i = 1, \dots, M$
\item for $k = 1, \dots, N$:
\begin{enumerate}\setlength{\itemsep}{1pt}
\item advance the control Kalman filter with $z_k$ to obtain
      $u_{k|k}$ and $P^u_{k|k}$
\item prediction: $T^i_{k|k-1} = g_{\theta^i}(T^i_{k-1|k-1})
      + B_{\theta^i} u_{k|k}$ for $i = 1, \dots, M$
\item compute the control sensitivities $B_{\theta^i} P^u_{k|k}
      B_{\theta^i}'$ ($\ell$ extra implicit solves per member)
\item assemble $\mathcal{P}_{k|k-1}\mathcal{H}'$ and
      $\mathcal{H}\mathcal{P}_{k|k-1}\mathcal{H}'$ from
      \eqref{eq:predcov}
\item analysis: update each $X^i = (\log\theta^i, T^i)$ with
      \eqref{eq:analysis}; project $\log\theta^i$ onto the prior box
\end{enumerate}
\end{enumerate}
\end{algorithm}

\begin{remark}
When $f_\theta \equiv 0$ and the boundary operator is Dirichlet,
$g_\theta$ is linear and Algorithm~\ref{alg:enmkf} reduces to the EnMKF
of \citet{Iglesias2018IP}.
\end{remark}

\subsection{Parameter-dependent observation operators}
\label{sec:condlinear}

In applications the observation operator often depends on the unknown
parameters.  The diffusive-flux observation of Section~\ref{sec:ex1} has
the form $h_\theta(T) = D_\theta H T$ with $D_\theta$ a diagonal matrix
whose entry for the flux is the unknown diffusivity $D$, so that this
component of the observation scales linearly with a parameter.
\citet{Iglesias2018IP} handled a parameter-scaled observation of this
kind by moving the scaling into the innovation while keeping a fixed
stencil in the gain.  Under the wide priors used here, that rescaling
couples the innovation noise to the spread of the parameter, and we
observed divergence of the parameter ensemble within a few assimilation
steps.

A consistent alternative is to augment the ensemble with the predicted
observations $G^i = h_{\theta^i}(T^i_{k|k-1})$ and apply the analysis
in the joint space, as in nonlinear-observation formulations of the
EnKF \citep{Evensen2009}.  The gain then uses the sample
cross-covariances of $(X, G)$, and the marginalization term of
\eqref{eq:predcov} is projected onto the observation coordinates
analytically: with $\Sigma_i = B_{\theta^i} P^u_{k|k} B_{\theta^i}' +
W_k$ and $J_i = D_{\theta^i} H$ the conditional Jacobian of the
observation map,
\begin{equation}
\label{eq:obsinflation}
\Delta P_{TG} = \frac{1}{M} \sum_{i=1}^M \Sigma_i J_i', \qquad
\Delta P_{GG} = \frac{1}{M} \sum_{i=1}^M J_i \Sigma_i J_i',
\end{equation}
added to the corresponding blocks of the sample covariance.  These two
terms are the Monte Carlo estimates of the posterior expectations
$\E_\zeta[\Sigma_\theta J_\theta']$ and $\E_\zeta[J_\theta \Sigma_\theta
J_\theta']$, that is, the projection of the marginalization block
$\Sigma_\theta$ onto the observation coordinates
through the conditional Jacobian $J_\theta = D_\theta H$.  The
construction is exact under the conditional linearity of $h_\theta$ in
$T$ and reduces to the previous section when $D_\theta = I$, so that
$J_\theta = H$ and $\Delta P_{GG}$, $\Delta P_{TG}$ recover the state
block of \eqref{eq:predcov}.

\subsection{Error-state EnMKF for second-order-in-time dynamics}
\label{sec:errorstate}

Per-step marginalization treats the filtered boundary uncertainty as
new at every step: the inflation $B_\theta P^u_{k|k} B_\theta'$ is
added at step $k$ regardless of how much of that same uncertainty was
already injected at step $k-1$.  For parabolic problems this
double-counting is mild, because the dynamics dissipate past boundary
influence and $\|B_\theta\|$ is uniformly bounded in $\Delta t$.  For
second-order-in-time problems it is not mild.  Writing the wave
equation as a first-order system in (displacement, velocity) and
discretizing with the trapezoidal rule, the sensitivity of the discrete
velocity to the current boundary displacement contains the factor
$2/\Delta t$: a boundary displacement error $\delta$ applied within one
step implies a velocity perturbation of order $\delta / \Delta t$.  If
the boundary error were white in time this would be the correct
physics.  It is not white: by \eqref{eq:errordyn} its correlation time
is governed by $\Phi_k$, and for smooth boundary signals it extends
over many steps.  Adding the full filtered variance independently at
every step therefore over-inflates the velocity block by a factor of
order $\tau_c / \Delta t$, where $\tau_c$ is the error decorrelation
time.  In the cable problem of Section~\ref{sec:ex3} this over-inflates the
near-boundary velocity block and makes per-step marginalization the
least accurate filter tested, with the largest damping bias; the
truncated prior nonetheless keeps the ensemble bounded, so the failure
is one of accuracy rather than outright divergence.

\begin{remark}
\label{rem:toy}
A scalar model isolates the mechanism.  Consider a displacement $w_k$
whose trapezoidal velocity update is $v_k = (2/\Delta t)(w_k - w_{k-1})
- v_{k-1}$, and suppose the boundary injects a displacement error $e_k$
into $w_k$, so that the boundary contribution to the velocity error is
$(2/\Delta t)(e_k - e_{k-1})$.  If $e_k$ is temporally persistent, then
$e_k \approx e_{k-1}$, the difference nearly cancels, and the velocity
error remains of order unity.  If instead $e_k$ is white in time with
variance $\sigma^2$, then $\mathrm{Var}(e_k - e_{k-1}) = 2\sigma^2$ and
the velocity-error variance is $8\sigma^2/\Delta t^2$, which diverges as
$\Delta t \to 0$.  Per-step marginalization and per-step resampling both
impose the white model, whereas the exact filtered error
\eqref{eq:errordyn} is persistent with correlation time $\tau_c$; the
spurious inflation is the ratio of $8\sigma^2/\Delta t^2$ to its
persistent counterpart, of order $\tau_c/\Delta t$.  Propagating $e_k$
with \eqref{eq:epsprop} restores the cancellation.
\end{remark}

The remedy keeps the marginalization philosophy, namely that everything
known in closed form about the boundary error should be used exactly,
but represents the error in time rather than step by step.  Each
ensemble member carries a realization $\varepsilon^i_k$ of the control
error state, propagated with the exact dynamics \eqref{eq:errordyn},
\begin{equation}
\label{eq:epsprop}
\varepsilon^i_k = \Phi_k\, \varepsilon^i_{k-1} + \eta^i_k, \qquad
\eta^i_k \sim \N(0, \Xi_k),
\end{equation}
and the member's boundary value is $u^i_k = u_{k|k} + H^u
\varepsilon^i_k$.  The matrices $\Phi_k$ and $\Xi_k$ come from the
control Kalman filter at no extra cost, so the temporal correlation of
the boundary error is exact, not modeled.  The error states join the
augmented vector in the analysis step, so the data update the joint
ensemble of parameters, states, and boundary errors, and the
cross-correlations between them are retained.  Model noise is sampled
as in a standard stochastic EnKF.  Algorithm~\ref{alg:enmkfeps}
summarizes the variant.

\begin{algorithm}[t]
\caption{Error-state EnMKF (temporally persistent boundary errors)}
\label{alg:enmkfeps}
\begin{enumerate}\setlength{\itemsep}{1pt}
\item draw $\theta^i \sim \rho_\Theta$, $T^i_0 \sim \N(T_0, P_0)$,
      $\varepsilon^i_0 = 0$, $i = 1, \dots, M$
\item for $k = 1, \dots, N$:
\begin{enumerate}\setlength{\itemsep}{1pt}
\item advance the control Kalman filter with $z_k$; store $u_{k|k}$,
      $\Phi_k$, $\Xi_k$
\item propagate the error states with \eqref{eq:epsprop} and set
      $u^i_k = u_{k|k} + H^u \varepsilon^i_k$
\item prediction: $T^i_{k|k-1} = g_{\theta^i}(T^i_{k-1|k-1})
      + B_{\theta^i} u^i_k + w^i_k$, $w^i_k \sim \N(0, W_k)$
\item analysis: stochastic EnKF update of the joint ensemble
      $(\log\theta^i, T^i, \varepsilon^i)$; project $\log\theta^i$
      onto the prior box
\end{enumerate}
\end{enumerate}
\end{algorithm}

The error states are initialized at $\varepsilon^i_0 = 0$, which assumes
no initial control error and is consistent with starting the
assimilation from the filtered mean of the control Kalman filter.  If a
nonzero initial covariance $P^u_{0|0}$ is available, the $\varepsilon^i_0$
should instead be drawn from the corresponding Gaussian in the companion
coordinates, so that the initial ensemble carries the known boundary
uncertainty.

The error-state filter is a sampling method, but it differs from the
modified EnKF baseline below in one essential respect:
the boundary-error realizations are consistent in time.  Resampling
$u^i_k \sim \N(u_{k|k}, P^u_{k|k})$ independently at every step makes
the boundary error white, which misrepresents its spectrum, suppresses
its low-frequency content, and biases the damping and stiffness
parameters that are identified precisely from that part of the
spectrum.  A practical rule for choosing between the two variants is to use
deterministic inflation (Algorithm~\ref{alg:enmkf}) for dissipative
dynamics with bounded control sensitivity, and the error-state filter
(Algorithm~\ref{alg:enmkfeps}) whenever the one-step control
sensitivity grows like $1/\Delta t$ or the boundary-error correlation
time is long compared with the assimilation interval.

\subsection{Baseline filters}
\label{sec:baselines}

Two baselines are used throughout Section~\ref{sec:examples}.  The
\emph{modified EnKF} samples $u^i_k \sim \N(u_{k|k}, P^u_{k|k})$
independently at each step, samples the model noise, and uses the plain
sample covariance in the analysis; it is the natural Monte Carlo
counterpart of the marginalized prediction and coincides with the
comparison filter of \citet{Iglesias2018IP}.  The \emph{noise-blind
EnKF} plugs the raw measurements $z_k$ into the boundary, which is the
common practice the EnMKF is designed to replace.

\section{Numerical examples}
\label{sec:examples}

Each example is a coefficient inverse problem: physical parameters of a
partial differential equation are identified from noisy interior
observations while the boundary data are themselves known only through
noisy records.  The boundary noise enters the forward model rather than
the likelihood, so it biases the parameter estimates in a way that
additional interior data do not remove; controlling this bias is the
purpose of the marginalization.  The two examples instantiate this
structure for a parabolic and a second-order-in-time problem.

All experiments use synthetic data generated on grids finer in space
and time than the assimilation grids, so the filters operate under
model discrepancy as they would with field data.  Reported uncertainty
is the ensemble standard deviation.  Implementations are in Python with
the per-member systems solved by a Thomas algorithm, vectorized across the
ensemble.

\subsection{Example 1: contaminant transport with second-order decay}
\label{sec:ex1}

The first example is a semilinear parabolic inverse problem from
contaminant transport, chosen to be self-contained for a general reader.
A solute diffuses and reacts in a one-dimensional column of length $L$,
\begin{equation}
\label{eq:contam}
\frac{\partial c}{\partial t}
 = D\,\frac{\partial^2 c}{\partial x^2} - k\,c^2,
 \qquad x \in (0, L),
\end{equation}
where the second-order (bimolecular) decay $-k c^2$ makes the dynamics
semilinear.  The left boundary exchanges with an external reservoir of
concentration $u(t)$ through a linear mass-transfer law, and the right
boundary is sealed,
\begin{equation}
\label{eq:contam_bc}
-D\,\frac{\partial c}{\partial x}(0,t)
 = h\,\bigl(u(t) - c(0,t)\bigr),
 \qquad \frac{\partial c}{\partial x}(L,t) = 0 .
\end{equation}
The reservoir concentration $u(t)$ is the control, known only through a
sensor record $z_k$ with measurement noise of standard deviation
$0.15$; this is the noisy-boundary setting the marginalization
addresses.  Both boundary operators are affine in $u$, so the
marginalization is exact and no approximation of the boundary term is
needed.  The unknown parameters are $\theta = (D, k, h)$: the
diffusivity, the reaction rate, and the mass-transfer coefficient, with
truth $(0.10, 1.00, 0.30)$ in the nondimensional units of
\eqref{eq:contam} and log-uniform priors $D \sim U(0.03, 0.30)$, $k \sim
U(0.30, 2.50)$, $h \sim U(0.05, 1.00)$.

The observations are the concentration at three interior stations, $x =
0.2, 0.4, 0.6$, and the diffusive flux $-D\,\partial_x c$ at $x = 0.3$.
The flux is linear in the state but scaled by the unknown diffusivity,
$h_\theta(c) = D\,H c$, a parameter-dependent observation of the form
treated in Section~\ref{sec:condlinear}: the ensemble is augmented with
the predicted observations and the marginalization term is projected
into the observation space.  Data are generated over $400$ steps of size
$\Delta t = 0.005$ on a grid of $80$ cells and assimilated on $40$, with
the diffusion advanced implicitly and the decay explicitly.

\begin{figure}[t]
\centering
\includegraphics[width=0.95\textwidth]{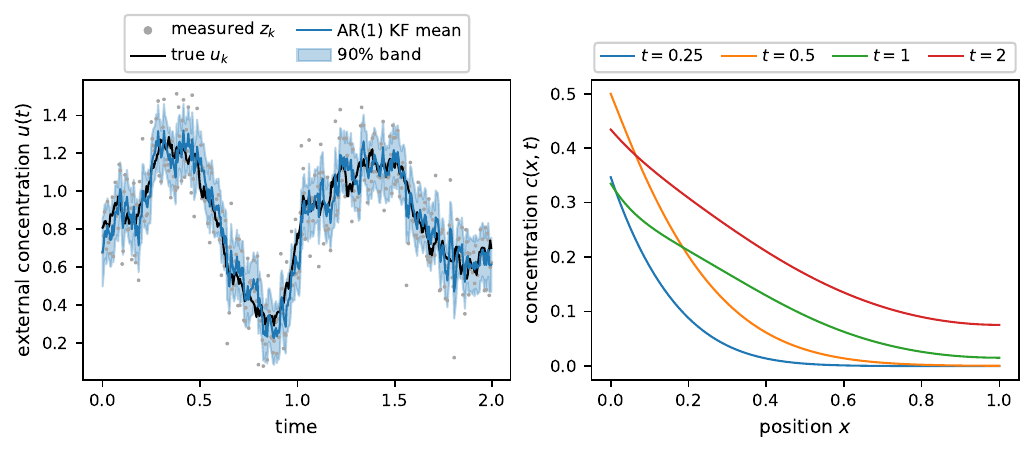}
\caption{Example 1, the noisy reservoir record with the AR(1) control
Kalman filter (left) and the concentration field decaying from the
mass-transfer boundary into the sealed column (right).}
\label{fig:ex1data}
\end{figure}

\begin{figure}[t]
\centering
\includegraphics[width=0.99\textwidth]{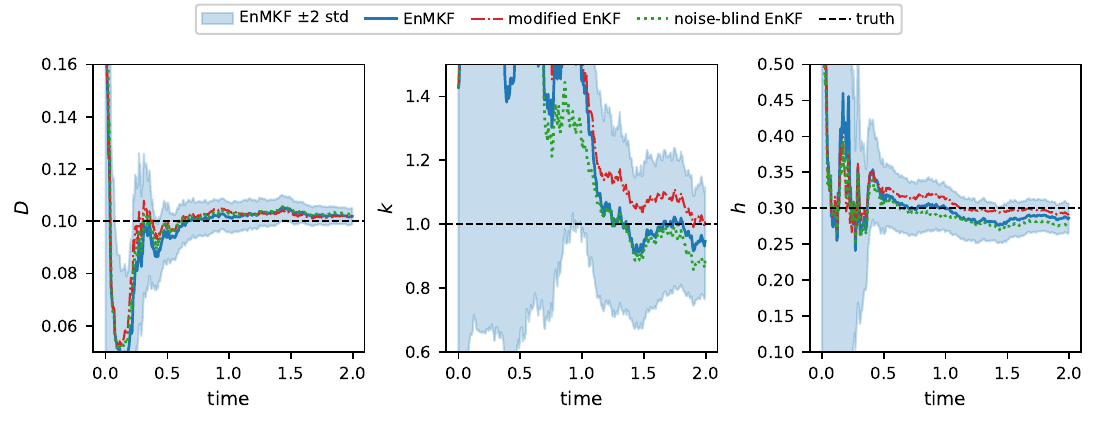}
\caption{Example 1, parameter trajectories at $M = 100$: generalized
EnMKF mean and two standard deviations (band), modified EnKF,
noise-blind EnKF, and truth.  The noise-blind filter attenuates the
transfer coefficient $h$ and the reaction rate $k$; the EnMKF recovers
all three parameters.}
\label{fig:ex1param}
\end{figure}

Figure~\ref{fig:ex1data} shows the reservoir record filtered by the
AR(1) control model and the decaying concentration field.
Figure~\ref{fig:ex1param} reports the identification with $M = 100$.
The generalized EnMKF recovers the three parameters, whereas the
noise-blind filter, which feeds the raw reservoir record into the
boundary, is biased: because the noisy reservoir concentration
multiplies the unknown transfer coefficient in \eqref{eq:contam_bc},
treating it as exact attenuates $h$ and, through the coupled dynamics,
pulls the reaction rate $k$ low.  This is an errors-in-variables bias;
it enters through the model rather than the likelihood, so it does not
diminish as more interior data are collected.

Table~\ref{tab:ex1} summarizes eight independent runs.  The EnMKF
recovers the diffusivity, reaction rate, and transfer coefficient to
within $1.7\%$ on average.  The modified EnKF, which samples the
reservoir uncertainty rather than marginalizing it, is comparably
accurate but noisier.  The noise-blind filter carries a persistent bias
of about $11\%$ in $k$ and $6\%$ in $h$.

\begin{table}[t]
\centering
\caption{Example 1, parameter estimates averaged over eight independent
runs at $M = 100$ (truth $D = 0.10$, $k = 1.00$, $h = 0.30$).  The
noise-blind filter attenuates $k$ and $h$; the marginalized filter
recovers all three.}
\label{tab:ex1}
\begin{tabular}{lccc}
\toprule
 & $D$ & $k$ & $h$ \\
\midrule
EnMKF            & 0.102 & 1.015 & 0.296 \\
modified EnKF    & 0.103 & 1.015 & 0.297 \\
noise-blind EnKF & 0.103 & 0.895 & 0.283 \\
\bottomrule
\end{tabular}
\end{table}

\begin{figure}[t]
\centering
\includegraphics[width=0.5\textwidth]{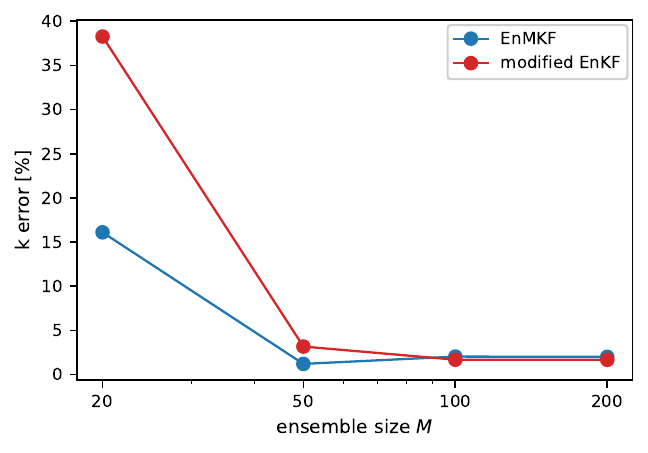}
\caption{Example 1, ensemble-size study: relative error in the reaction
rate $k$ against the ensemble size $M$, for the generalized EnMKF and
the modified EnKF (mean over six runs).}
\label{fig:ex1mstudy}
\end{figure}

Figure~\ref{fig:ex1mstudy} repeats the identification across ensemble
sizes.  The marginalized filter attains its small-sample accuracy by
about $M = 50$, where the modified EnKF still carries more than twice the
error in $k$ and a wider run-to-run scatter; the two agree once the
ensemble is large enough to resolve the sampled boundary uncertainty.
This is the ensemble-size economy documented for the linear filter,
preserved in the semilinear setting.

\subsection{Example 2: nonlinear cable vibration with noisy support
motion}
\label{sec:ex3}

The second example is hyperbolic and exercises the error-state filter of
Section~\ref{sec:errorstate}.  A tensioned cable of length $10$ m on a
cubic elastic foundation is driven at one end by support motion, as in
a shake-table experiment,
\begin{equation}
\label{eq:cable}
\frac{\partial^2 w}{\partial t^2} + 2\gamma\, \frac{\partial w}{\partial t}
 = c^2\, \frac{\partial^2 w}{\partial x^2} - \kappa\, w^3,
\qquad w(0, t) = u(t), \qquad w_x(10, t) = 0,
\end{equation}
with $\theta = (c, \gamma, \kappa)$, truth $(20\;\mathrm{m/s},\;
0.8\;\mathrm{s^{-1}},\; 5 \times 10^3\;\mathrm{m^{-2}s^{-2}})$ and
log-uniform priors on $(10, 35)$, $(0.1, 3)$, and $(5 \times 10^2,
2 \times 10^4)$.  The support motion is a filtered-noise record of the
Kanai--Tajimi type \citep{Kanai1957} with a peak of $50$ mm (cable response up to $64$ mm),
known only through a displacement record with $5$ mm noise, the level
expected when the support displacement is reconstructed from
accelerometers.  Displacements are observed at $x = 2.5, 5, 7.5, 10$ m
with $0.5$ mm noise at $\Delta t = 2$ ms over six seconds.  The truth
uses 200 cells and four sub-steps per observation interval; the filters
use 50 cells.  The trapezoidal discretization conserves energy in the
undamped limit, so the damping estimate is not contaminated by
numerical dissipation.

Per-step marginalization behaves on this problem in the way predicted in
Section~\ref{sec:errorstate}.  The velocity's sensitivity to the boundary displacement scales like
$2/\Delta t$, so it turns the few-millimeter filtered uncertainty into a
large per-step inflation of the near-boundary velocity, which the gain absorbs as spurious high-frequency energy.  The
consequence, quantified below, is the largest damping bias and the least
accurate displacement reconstruction of any filter tested; the truncated
prior keeps the ensemble bounded, so the failure is one of accuracy
rather than outright divergence.  The error-state filter, which
propagates the boundary error with its exact temporal correlation, has
no such difficulty.

\begin{figure}[t]
\centering
\includegraphics[width=0.99\textwidth]{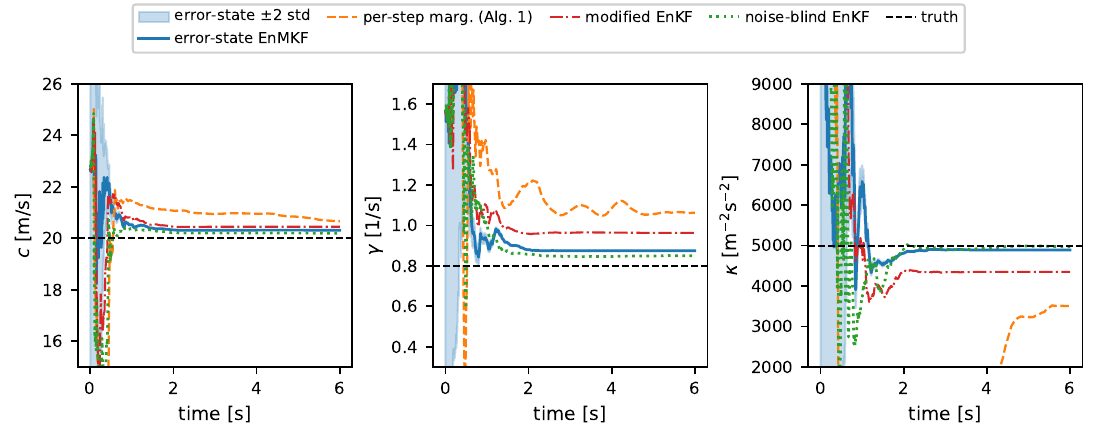}
\caption{Example 2, parameter trajectories: error-state EnMKF mean and
two standard deviations (band), per-step marginalization (Algorithm~1),
modified EnKF, noise-blind EnKF, and truth.  Per-step marginalization
and the modified EnKF both over-damp because they treat the boundary
error as white in time; the error-state EnMKF recovers the damping and
the cubic stiffness.}
\label{fig:ex3param}
\end{figure}

\begin{figure}[t]
\centering
\includegraphics[width=0.9\textwidth]{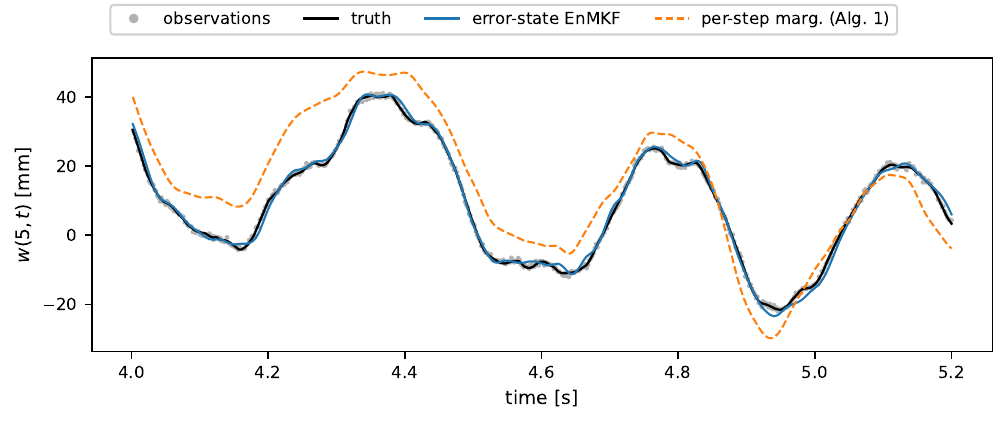}
\caption{Example 2, displacement at $x = 5$ m between $t = 4$ and $5.2$
s: observations, truth, error-state EnMKF mean, and per-step
marginalization (Algorithm~1) mean, which is visibly noisier.}
\label{fig:ex3state}
\end{figure}

Figure~\ref{fig:ex3param} shows the identification results with $M =
100$.  The error-state EnMKF estimates $c = 20.3$, $\gamma = 0.87$,
$\kappa = 4.89 \times 10^3$.  Per-step marginalization (Algorithm~1),
which adds the boundary variance as a white inflation of the prediction
covariance, over-damps the most, $\gamma = 1.06$, because the $2/\Delta
t$ velocity sensitivity makes that inflation largest on the velocity
block.  The modified EnKF, whose only difference from the error-state
filter is the white-in-time resampling of the boundary values,
overestimates the damping by a similar margin ($\gamma = 0.96$) and
underestimates the cubic stiffness by $13\%$ ($\kappa = 4.35 \times
10^3$).  The noise-blind filter lands closer to the truth on $\gamma$
and $\kappa$ for this realization, because feeding the raw noisy record
acts as unintended additive inflation, but it pays in state accuracy.
Table~\ref{tab:ex3rmse} reports the root-mean-square error of the
reconstructed displacements at the four sensors over the final four
seconds: the error-state EnMKF is the most accurate at every station and
per-step marginalization the least, with the two white-boundary filters
in between.  All filters remain overconfident relative to the residual
discretization discrepancy in $\gamma$, a limitation discussed in
Section~\ref{sec:conclusions}.  Figure~\ref{fig:ex3state} shows a
segment of the reconstructed displacement at mid-span.

\begin{table}[t]
\centering
\caption{Example 2, displacement RMSE in millimeters at the four
sensors, computed over $t \in [2, 6]$ s for the representative run of
Figure~\ref{fig:ex3param}.}
\label{tab:ex3rmse}
\begin{tabular}{lcccc}
\toprule
 & $x = 2.5$ m & $x = 5$ m & $x = 7.5$ m & $x = 10$ m \\
\midrule
error-state EnMKF & 0.91 & 1.07 & 1.78 & 1.33 \\
per-step marginalization (Alg.~1) & 5.22 & 7.17 & 8.96 & 10.29 \\
modified EnKF     & 1.55 & 1.38 & 2.03 & 1.91 \\
noise-blind EnKF  & 2.33 & 1.74 & 2.00 & 2.44 \\
\bottomrule
\end{tabular}
\end{table}

The single realization of Figure~\ref{fig:ex3param} and
Table~\ref{tab:ex3rmse} is representative, but the comparison across
realizations is the more telling one, because the two baseline filters
are not merely biased: they occasionally fail.
Table~\ref{tab:ex3robust} summarizes sixteen independent runs.  The
error-state EnMKF is accurate and stable: its median damping and
cubic-stiffness errors are $11\%$ and $5\%$, its worst-case damping and
cubic-stiffness errors over the sixteen runs are $15\%$ and $10\%$, and
no run is driven to a prior bound.  Per-step marginalization carries the largest damping bias of the
four filters (median $25\%$), consistent with the over-inflated velocity
block, but it does not diverge on any run.  The modified EnKF carries a
comparable damping bias (median $22\%$) for the related reason that the
whitened boundary error injects spurious high-frequency energy that the
filter absorbs as extra damping.  The noise-blind filter has a
deceptively good median, since feeding the raw record acts as an
unintended inflation that is relatively small on typical realizations; but it has
no control on the tail, with a worst-case cubic-stiffness error near
$300\%$ and one of the sixteen runs collapsing onto the prior bounds.
Its mean error is therefore dominated by rare catastrophic runs
and is not reproducible across independent trials, which is why we report the
median together with the worst case rather than a mean over a handful of
runs; this summary is faithful to a heavy-tailed error distribution.
The displacement reconstruction separates the filters most clearly: the
median sensor RMSE of the error-state EnMKF is $0.9$, $1.1$, $1.8$, $1.4$
mm at the four stations, against $1.7$, $1.6$, $2.2$, $2.3$ mm for the
modified EnKF, $2.4$, $2.0$, $2.3$, $2.7$ mm for the noise-blind filter,
and $4.2$, $4.1$, $3.8$, $5.1$ mm for per-step marginalization, so the
error-state filter is the most accurate at every station and per-step
marginalization the least.

\begin{table}[t]
\centering
\caption{Example 2, robustness over $N = 16$ independent runs at $M = 100$, with truth $c = 20$ m/s, $\gamma = 0.8$
s$^{-1}$, and $\kappa = 5 \times 10^3$ m$^{-2}$s$^{-2}$.  For each
parameter we report the median and, in parentheses, the worst-case
relative error over the runs, in percent; the last column counts the
runs in which any parameter was driven to a prior bound.}
\label{tab:ex3robust}
\begin{tabular}{lcccc}
\toprule
 & $c$ & $\gamma$ & $\kappa$ & diverged \\
\midrule
error-state EnMKF & 1.9 (2.4)  & 10.7 (15.1) & 4.8 (10.1)  & 0/16 \\
per-step marginalization (Alg.~1) & 2.7 (5.9) & 25.1 (59.4) & 2.8 (64.3) & 0/16 \\
modified EnKF     & 2.8 (4.9)  & 22.0 (43.7) & 12.5 (70.4) & 0/16 \\
noise-blind EnKF  & 1.1 (14.5) & 6.6 (87.2)  & 3.7 (296.5) & 1/16 \\
\bottomrule
\end{tabular}
\end{table}

\section{Summary and conclusions}
\label{sec:conclusions}

The ensemble-marginalized Kalman filter extends to semilinear PDEs with
little structural change, because the property it relies on is the
affine entry of the boundary data into the discrete dynamics, not the
linearity of the dynamics themselves.  IMEX discretizations supply this
property by construction.
One extension widens the reach of this parabolic filter:
parameter-dependent observation operators are handled consistently by
augmenting the ensemble with predicted observations and projecting the
marginalization term onto the observation space.  Boundary operators in
which the control itself enters nonlinearly, such as radiative exchange,
leave the exact-marginalization framework; they can be accommodated
approximately by a delta-method linearization of the control entry, an
extension we leave to future work.

The wave example highlights the main practical lesson of the paper. Per-step
marginalization implicitly treats the
filtered boundary error as white in time.  For dissipative dynamics
with bounded control sensitivity this approximation is mild, and the
deterministic-inflation filter retains the advantages documented in the
linear case: no ensemble collapse, smaller bias at small ensemble
sizes, and a quarter to a half of the ensemble required by a sampled
treatment for the same accuracy.  For second-order-in-time dynamics the
one-step control sensitivity of the velocity grows like the inverse
time step and the whiteness assumption fails by a factor of the error
correlation time over the time step.  The error-state variant removes
the assumption at negligible cost by propagating boundary-error
realizations with the exact innovations-form dynamics of the control
Kalman filter, and it reduced both the parameter bias and the state
reconstruction error relative to the resampling baseline.

Three limitations deserve note.  First, the filters inherit the usual
caveat of ensemble Kalman methods: the analysis step is exact only
under Gaussianity, and the parameter estimates are not posterior
samples in the strict Bayesian sense \citep{ErnstSprungk2015};
the sampling error that accumulates through the repeated analysis steps
is not tracked here.  Second,
under model discrepancy between the assimilation grid and reality, all
filters tested here are overconfident, because the ensemble spread
measures sampling uncertainty given the model rather than structural
error.  The noise-blind filter is additionally biased, as Example~1 shows, so
its already optimistic intervals are mis-centred; the marginalized
filter removes that bias but remains, like all the ensemble filters
here, somewhat overconfident.  The
marginalization therefore mitigates the overconfidence but does not cure
it, and joint estimation of a discrepancy term or of the noise
hyperparameters is a natural extension.  Third, the
conditional-independence assumption of
Proposition~\ref{prop:marg} is removed by the error-state filter only
for the boundary error; correlations between model noise and boundary
noise, which arise when both originate from the same unresolved
physics, remain outside the present framework.

The control formulation is not limited to boundary data.  Uncertain
source amplitudes, inflow hydrographs, and reconstructed excitation
records all fit the companion model \eqref{eq:control}, so the methods
apply to flood routing, pollutant source problems, and structural
identification under base excitation, in addition to the building
physics application that motivated the original filter.

\bibliography{refs}

\end{document}